\documentclass[twocolumn,showpacs,preprintnumbers]{revtex4}
\usepackage{dcolumn}
\usepackage{bm}
\input epsf

\begin{document}

\preprint{physics/08MMNNN}

\title{A chilean seismic regionalization through a Kohonen neural network}

\author{Jorge Reyes$^{1}$}
\email{daneel@geofisica.cl}

\author{V\'{\i}ctor H. C\'{a}rdenas$^{2}$}
\email{victor@dfa.uv.cl}

\affiliation{$^{1}$TGT, San Renato 217, Los Andes, Chile}

\affiliation{$^{2}$Departamento de F\'isica y Astronom\'ia,
Universidad de Valpara\'iso, Valpara\'iso, Chile}

\begin{abstract}
A study of seismic regionalization for central Chile based on a
neural network is presented. A scenario with six seismic regions is
obtained, independently of the size of the neighborhood or the reach
of the correlation between the cells of the grid. The high
correlation between the spatial distribution of the seismic zones
and geographical data confirm our election of the training vectors
of the neural network.
\end{abstract}


\maketitle

%%%%%%%%%%%%%%%%%%%%%%%%%%%%%%%%%%%%%%%%%%%%%%%%%%%%%%%%%%%%%%%%%%%%%%
\section{Introduction}

The idea of seismic regionalization can be traced back to 1941 when
Gorshkov \cite{urss41} published one of the first studies on seismic
regionalization for the URSS. Based on these earlier studies,
Richter \cite{richter}, in a paper entitled ``Seismic
regionalization'', founded the basis of a methodology that allows to
have information dealing with preventing the potential damage that
an earthquake may cause through the definition of hazard zones.

In a seismic country like Chile, and based on the urgent goals
established by Richter, a proper regionalization map is desirable.
Attempts to fill the gap in these matters have been done from time
to time. In \cite{GL58} is found probably the first regionalization
study for Chile where the authors Gajardo and Lomnitz used the same
methods of seismic correlation Tsuboi used in the case of Japan.
Later works on the subject were mainly focused on engineering
aspects as Walkner \cite{W64}, Labbe \cite{L76} and Barrientos
\cite{barrientos}, discussing on seismic risk. Essentially, these
studies used the technique proposed by Cornell \cite{C72} and
Algermissen and Perkings \cite{AyP76}. A more recent study was made
by Martin \cite{amartin}. Here the author used a longitudinal
division criteria for the country using the line that separates the
deep and superficial earthquakes, which leads to a high degree of
coupling between the Nazca and Southamerican plate. This criterium
enable him to identify two macro-zones in the continental zone: In
the coast, with hypocenters with depths lesser than $40$km, and the
mountain macro-zone. As a criteria for latitudinal division, Martin
computed the $b$ parameter of the Gutenberg-Richter law for an
initial surface, and then he varied the surface iteratively. If $b$
changes dramatically, it means that a seismic zone has been crossed.
These are the nine regions Martin found: in the coast; three zones.
In the deep mountain zone, he distinguished four zones. In the
surface mountain zone, he distinguished a single zone. Finally, the
method determined the aseismic zone of Magallanes.

In this work we implement a Kohonen neural network (from here on NN)
to determine the seismic zones. In general, a neural network
\cite{NN} can be defined as a computing system made up of a number
of simple, highly interconnected processing elements, which
processes information by their dynamic state response to external
inputs. NNs are typically organized in layers. Layers are made up of
a number of interconnected `nodes' which contain an `activation
function'. Patterns are presented to the network via the `input
layer', which communicates one or more `hidden layers' where the
actual processing is done through a system of weighted
`connections'. The hidden layers then link to an `output layer'
where the answer is output. Most NNs contain certain types of
`learning rule' which modify the weights of the connections
according to the input patterns which are exposed to.

In this work we use a special NN appropriated to discriminate
spatial distribution. A Kohonen NN \cite{KNN}(or as it is usually
called a self organising map) is an artificial neural network that
is trained using unsupervised learning to produce a low-dimensional
(typically two dimensional), discretized representation of the input
space of the training samples, called a map. The network must be fed
a large number of example vectors that represent, as close as
possible, the kinds of vectors expected during mapping. The examples
are usually provided several times. In this way, the NN finds by
itself the best distribution of zones, based on the training
vectors, without any other specification than geographical
characteristics.

In the context of seismic hazard or seismic risk, there have been
many attempts to use this technique. For example, in \cite{DTA90}
the authors used artificial NN to discriminate between earthquakes
and underground nuclear explosions. Also for seismic detection
\cite{WT95}, in which the NN is trained to recognize signal
patterns, and also for earthquake prediction \cite{GPR00}. This work
is organized as follows. The next section discusses the election of
the training vectors or training set, from which the NN starts its
learning process. Then, in section III we discusses the way we
perform the training process. Section IV is devoted to the analysis
of the results and we finish the paper with conclusions.

%%%%%%%%%%%%%%%%%%%%%%%%%%%%%%%%%%%%%%%%%%%%%%%%%%%%%%%%%%%%%%%%%%%%%%
\section{Training set}

The election of the training set is the crucial step towards the
implementation of a useful NN. The usefulness of a NN depends on
what we are interested to get from it. As we have discussed in the
previous section, the necessity is mainly the one of a hazard
categorizer in the Chilean territory. In this section we discuss the
criteria in the election of our training set.

\subsection{Database}

The idea was to associate the earthquake information by zone,
frequency and magnitude. We expected the NN finds by itself what we
call here `zone', by introducing in it specific information about
its seismic history. In this work we use as a database all the
earthquakes classified with Richter magnitude equal or larger than
$4.5$ Ms in the interval 1957 to 2007. The sources for the data we
used, were obtained from the catalog CERESIS \cite{ceresis} and from
the database of the USGS \cite{usas} (from 1973 to the present).

The area covered by our data set is Chile's continental territory
included inside a rectangle defined around the meridian $72^{\circ}$
W as the North-South axis, between the parallels $17^{\circ}$ S and
$56^{\circ}$ S (see the figures for details).

The interval of fifty years ensures us to have a very representative
set of data. It is long enough to have earthquake frequency
histograms with representative information which could be projected
to a larger period of time.

\subsection{The Kohonen training vectors}

In order to classify the zones, we divided the area selected in the
previous section by $156$ rectangular cells of $1^{\circ} \times
1^{\circ}$. Each cell was characterized by a seven dimensional
vector, indicating the observed historical seismicity between 1957
to 2007. The $7$D vector is explained below with more details:

  $x_1$: Mean deep of the seismic sources

  $x_2$: Number of earthquakes with magnitude $\geq 4.5$ Ms

  $x_3$: Number of earthquakes with magnitude $\geq 5.5$ Ms

  $x_4$: Number of earthquakes with magnitude $\geq 6.5$ Ms

  $x_5$: Maximum observed magnitude

  $x_6$: Central horizontal coordinate of the cell

  $x_7$: Central vertical coordinate of the cell

\vspace{0.2cm}

These are considered the minimum number of sensitive parameters
which characterize each event. No geophysical input was necessary.

%%%%%%%%%%%%%%%%%%%%%%%%%%%%%%%%%%%%%%%%%%%%%%%%%%%%%%%%%%%%%%%%%%%%%%
\section{Training Procedure}

The $156$ vectors were introduced into the Kohonen NN. The free
parameters in the running are two: the number of iterations and the
number of classes or regions. By inspection, it was observed that in
all the runnings, and after an average of $2500$ iterations, the NN
reached a stable configuration, indicating that the system had found
a local minimum.

\begin{figure}[tb]
\centering \leavevmode\epsfysize=10cm \epsfbox{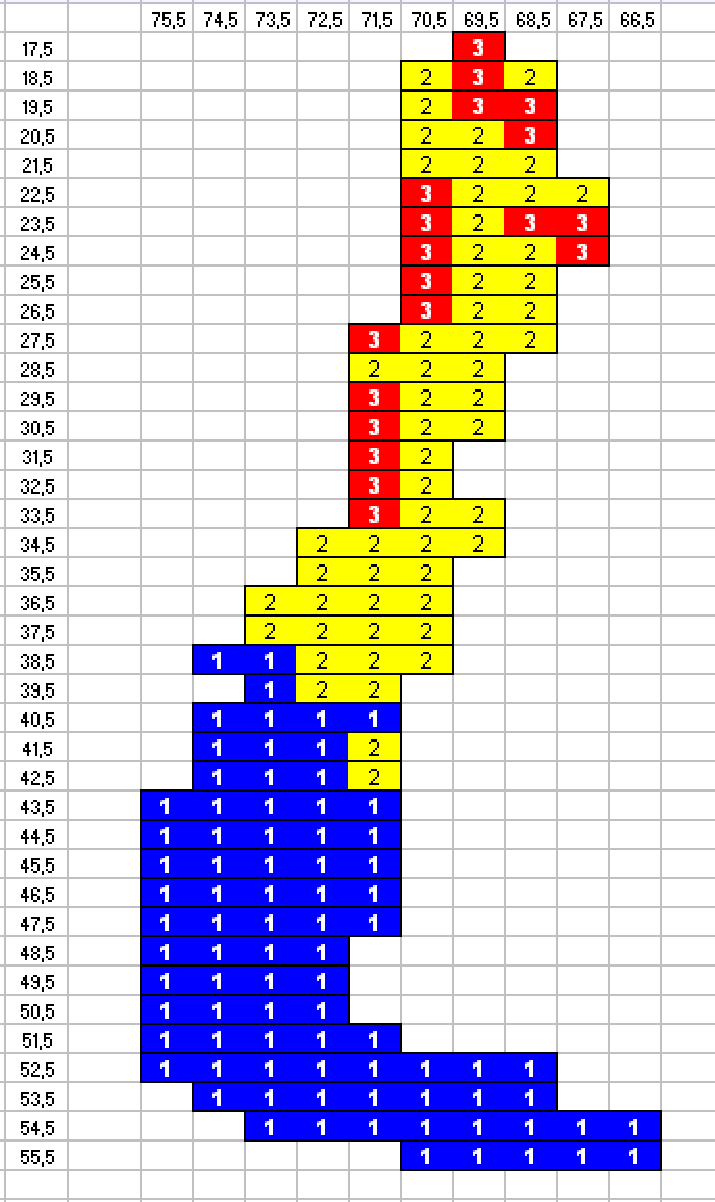}\\
\caption{\label{fig1} Meta stable configuration after 1000
 iterations of the neural network. This result was obtained independent
 of how many classes we ask for, revealing a strong .}
\end{figure}

In an initial training session, we forced the NN to create $5$, $10$
and $15$ classes. In all these cases, it was observed that during
the first $1000$ iterations, the system reached a stable
configuration with just three classes, as it is shown in
Fig.(\ref{fig1}).

We called these three classes ``the fundamental classes''. Let us
explain in detail each one of these. According to Fig. (\ref{fig1}),
in the northern part of the country, we distinguish two classes or
two seismic zones of high seismicity, associated to the interaction
between the Nazca and South-American plate. In the south, we
distinguish another class or seismic zone, with low seismicity,
associated to the interaction among the Southamerican, Scotia and
Antartic plates.

\begin{figure}[tb]
\centering \leavevmode\epsfysize=12cm \epsfbox{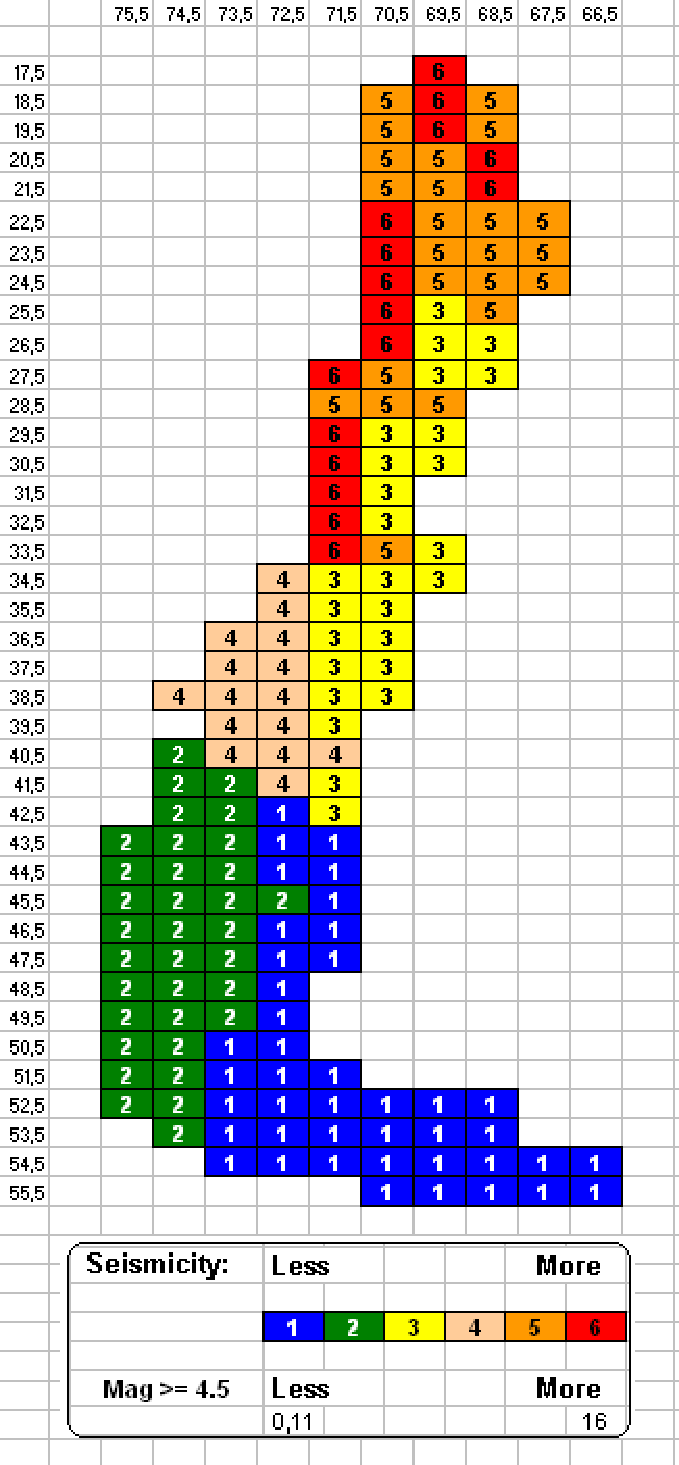}\\
\caption{\label{fig2} Transient stable configuration after 2500
 iterations and independent of the initial reach of the
 correlation (or initial size of the neighborhood). The numbers are explained
 in Table I.}
\end{figure}

Enhancing the detailed study, we forced the NN to catalog six
classes or seismic regions. In this case, after 2500 iterations, the
final result was the same, essentially independent of the initial
size of the neighborhood. Figure (\ref{fig2}) shows the final result
using 6 clusters or categories, 2500 iterations and an initial size
of the neighborhood of 5.

%%%%%%%%%%%%%%%%%%%%%%%%%%%%%%%%%%%%%%%%%%%%%%%%%%%%%%%%%%%%%%%%%%%%%%
\section{Analysis}

\subsection{Island regions}

\begin{figure}[htb]
\centering \leavevmode\epsfysize=9cm \epsfbox{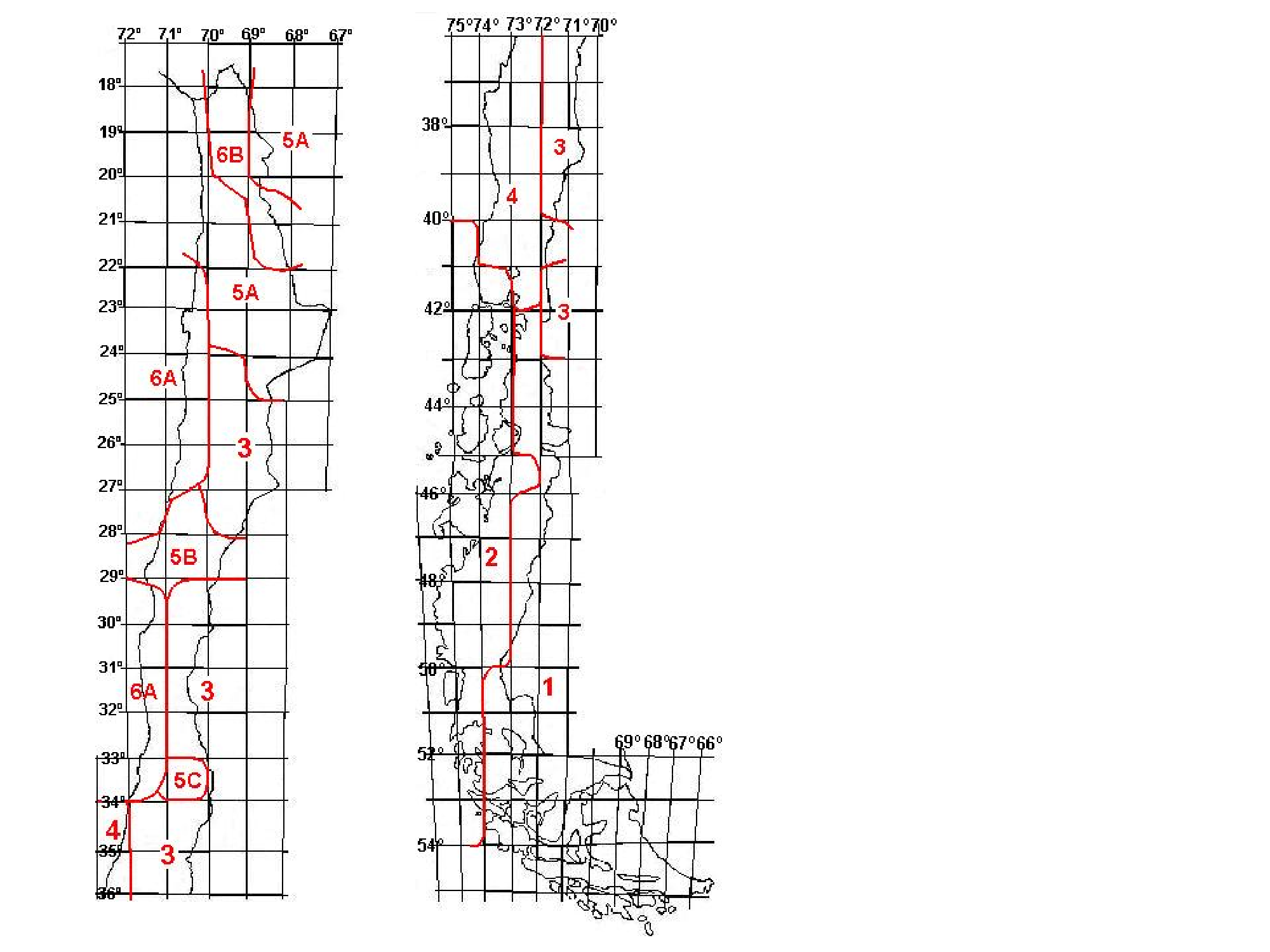}\\
\caption{\label{fig3} North and South part of Chile. Here is the
division of discussed seismic zones in the text.}
\end{figure}

As we can see in Fig.(\ref{fig3}), in certain regions some
discontinuities occurred, through an interruption of another region.
We called these island regions.

In the north part, region 6 becomes separated by two subregions;
region 6A (coast) and region 6B (mountain).

The north part, region 5 becomes separated in three subregions;
region 5A (including Iquique city), region 5B (including Copiap\'o
city) and region 5C (centered in Santiago city).

Although they are discontinuous regions, the Kohonen NN associates
these in the same category because they presented similar
seismicity. For that reason we stress to name it just region 5 and
6. Of course, the seismicity associated to each region depends on
the chosen vectors representing each cell.

\subsection{Gutenberg-Richter Law}

For each region we found the parameters of the Gutenberg-Richter
(GR) law. Here $N(\geq M)$ is the mean amount of earthquakes of
magnitude larger than $M$ observed in a year, in a normalized area
(equivalent to a cell of $1^\circ \times 1^\circ$ in the equator).

\begin{table}
\caption{\label{tab:table1}This is a summary of the results for the
parameters in the Gutenberg-Richter law. The equation reads $\log N
= a+bM$, where $M$ is the magnitude in Ms units. It also is given
the $\chi^2$ of the fit and the maximum magnitude (in Ms units) for
each region.}
\begin{ruledtabular}
\begin{tabular}{cccccc}
Region & a & b & $\chi^2$ & Max.Mag. (Ms)& $N(\geq4.5)$ \\
\hline
1 & 3.844 & -1.069 & 0.9073 & 5.5 & 0.11\\
2 & 4.209 & -0.922 & 0.8650 & 6.9 & 1.15 \\
3 & 3.298 & -0.699 & 0.9373 & 7.7 & 1.42\\
4 & 2.702 & -0.520 & 0.9721 & 8.3 & 2.30\\
5 & 3.784 & -0.685 & 0.9405 & 7.1 & 5.02\\
6 & 3.702 & -0.555 & 0.9388 & 7.8 & 16.03\\
\end{tabular}
\end{ruledtabular}
\end{table}

%%%%%%%%%%%%%%%%%%%%%%%%%%%%%%%%%%%%%%%%%%%%%%%%%%%%%%%%%%%%%%%%%%%%%%
\section{Conclusions}

Considering the semi-logarithmic GR law slope, we obtain the
hierarchy of the six regions, as it is shown in Table II.
\begin{table}{}
\caption{Here it is displayed the seismic zones under the hierarchy
of a growing order according to the $\left|b\right|$ parameter of
the Gutenberg-Richter Law. NN used as input data the amount of
seisms observed in the last fifty years, we assume that the final
hierarchy must be mainly based on the number of earthquakes which
are equal or larger than 4.5 Ms, according to Table I.}
\begin{ruledtabular}
\begin{tabular}{ccc}
Hierarchy & Zone & $\left|b\right|$ \\
\hline
1 & 4 & 0.520\\
2 & 6 & 0.555 \\
3 & 5 & 0.685 \\
4 & 3 & 0.699 \\
5 & 2 & 0.922 \\
6 & 1 & 1.069 \\
\end{tabular}
\end{ruledtabular}
\end{table}

In Table I, and looking at the first and last columns, we observe
the hierarchy based on the seismicity of the zones. As it was
explained in section III, the NN was trained with information about
the number of earthquakes per range. Then, the definition of
seismicity deduced from the NN must be associated to the magnitude
and angular orientation of the 7D vectors, which  strongly depends
on the number of earthquakes equal or larger than $4.5$ Ms.

In this context, zone 1 is an aseismic zone, and zone 6 is the most
seismic one.

It is possible that zones 5A and 6B are parts of a larger one
extending themselves towards Peru and Bolivia.

The zones 5B and 5C have an interruption between latitudes
$29^\circ$ and $32^\circ$, which coincides with a geographic
accident; the intermediate depression despair, being replaced by the
fusion between the Andes and coast mountains. Then, zone 5 is
strongly associated to the intermediate depression and zone 3 is
strongly related to the Andes mountains.

%\begin{figure}[htb]
%\centering \leavevmode\epsfysize=7cm \epsfbox{Fig3B.eps}\\
%\caption{\label{fig3b} South part of Chile. The same as the panel
%above.}
%\end{figure}

In the north part of the country, the high seismicity observed can
be understood as the result of the activity of the subduction
between the Nazca and Southamerican plates (at a rate of convergence
of $\sim 10$cm per year, as it is well known).

In the south, we obtain a low seismicity in agreement with the well
known result of a convergence of the Southamerican, Scotia and
Antartic plates, at a rate of $\sim 2$ cm per year.

In this work we have demonstrated that using a Kohonen NN, and the
database available up to date, it can be possible to differentiate
six seismic zones for continental Chilean territory. Some of them
are apparently disconnected zones (we called these here island
zones), but this is in fact an effect produced by considering just
an arbitrary portion of territory (that of the political Chilean
one) in the analysis. Also, the division between these island zones
can be clearly associated with major geographical accidents (as the
intermediate depression and some deep valleys) that were not used as
input data in the NN. All these facts show the power of using NN
analysis in seismic regionalization.

%%%%%%%%%%%%%%%%%%%%%%%%%%%%%%%%%%%%%%%%%%%%%%%%%%%%%%%%%%%%%%%%%%%%%%
\section*{Acknowledgments}

JR wants to thank TGT for the support through grant number 2122. VHC
wants to thank Rafael Valdivia for useful discussions.

%%%%%%%%%%%%%%%%%%%%%%%%%%%%%%%%%%%%%%%%%%%%%%%%%%%%%%%%%%%%%%%%%%%%%%

\end{document}